\PassOptionsToPackage{unicode}{hyperref}
\PassOptionsToPackage{dvipsnames,svgnames*,x11names*}{xcolor}

\documentclass[]{article}

\usepackage{arxiv}

\usepackage[utf8]{inputenc} 
\usepackage[T1]{fontenc}    
\usepackage{lmodern}        
\usepackage{xcolor}
\usepackage{hyperref}       
\usepackage{url}            
\usepackage{booktabs}       
\usepackage{amsfonts}       
\usepackage{nicefrac}       
\usepackage{microtype}      
\usepackage{graphicx}
\usepackage{setspace}
\usepackage{orcidlink}
\usepackage{fontspec}

\hypersetup{
  pdftitle={G-computation for causal effect estimation from observational hierarchical data with unmeasured cluster context},
  pdfkeywords={causal inference, g-computation, hierarchical data, unmeasured confounding, treatment effect heterogeneity, MICS},
  breaklinks=true,
  bookmarks=true,
  colorlinks=true,
  linkcolor={magenta},
  filecolor={Maroon},
  citecolor={blue},
  urlcolor={Blue},
  pdfcreator={LaTeX via pandoc}}

\title{G-computation for causal effect estimation from observational hierarchical
data with unmeasured cluster context}

\author{
    Shafayet Khan Shafee%
  \hspace{0.12em}\orcidlink{0009-0002-8021-3788}%
  \thanks{corresponding author}%
  \\
  Independent Researcher \\
  Dhaka, Bangladesh \\
  \texttt{\href{mailto:sshafee@isrt.ac.bd}{\nolinkurl{sshafee@isrt.ac.bd}}} \And
  Bishal Sarker%
  \hspace{0.12em}\orcidlink{0009-0001-0812-4142}%
  \thanks{Joint-second authors contributing equally (alphabetic ordering)}%
  \\
  Independent Researcher \\
  Dhaka, Bangladesh \\
  \texttt{\href{mailto:bsarker1@isrt.ac.bd}{\nolinkurl{bsarker1@isrt.ac.bd}}} \And
  Md. Niamul Islam Sium%
  \hspace{0.12em}\orcidlink{0009-0005-1822-8294}%
  \footnotemark[\value{footnote}]%
  \\
  University of Texas at El Paso \\
  El Paso, Texas, USA \\
  \texttt{\href{mailto:msium@miners.utep.edu}{\nolinkurl{msium@miners.utep.edu}}}}

\usepackage[round,sort]{natbib}

\usepackage{float}
\usepackage{caption}
\DeclareCaptionFont{tenpt}{\fontsize{9.5}{11.5}\selectfont}
\usepackage{mathtools}
\usepackage{booktabs}
\usepackage{array}
\usepackage{doi}
\usepackage{url}
\usepackage{multirow}
\usepackage{threeparttable}
\begin{document}
\maketitle

\begin{abstract}
Observational studies frequently involve hierarchical data structures in which
individuals are nested within higher-level units. In such settings, unmeasured
cluster-level factors may confound the treatment-outcome relationship and may
additionally induce treatment effect heterogeneity across clusters, complicating
causal effect estimation. We formalize the use of g-computation for hierarchical
observational data by incorporating random-effects models (REM) as outcome
models and propose a within-group g-computation strategy designed to reduce
bias arising from unmeasured cluster context. The approach groups clusters
according to their observed treatment prevalence and performs g-computation
within groups before aggregating group-specific estimates. Through extensive
Monte Carlo simulations, we compare the standard and within-group implementations
of g-computation using both linear models and REM. Results show that both
standard and within-group REM-based implementations substantially reduce bias
when the unmeasured cluster-level variable acts solely as a confounder, whereas
the proposed within-group REM estimator achieves the lowest RMSE when the
unmeasured cluster-level factor acts as both a confounder and a source of
treatment effect heterogeneity. We apply the proposed within-group REM estimator
to estimate the causal effect of adolescent pregnancy on the child height-for-age
Z-score using 2019 Bangladesh MICS data, obtaining an estimated effect of -0.12
(95\% bootstrap CI: {[}-0.18, -0.06{]}). The proposed within-group g-computation
framework offers a strategy for reducing bias from unmeasured cluster-level
confounding and treatment effect heterogeneity in hierarchical observational
studies.
\end{abstract}


\setstretch{1.05}

\section{Introduction}\label{sec-01}

Causal inference seeks to estimate the effect of a treatment or exposure (used
interchangeably hereafter) on an outcome by comparing potential outcomes under
different treatment conditions for the same unit. As formalized by Rubin, the
individual treatment effect for unit \(i\) is defined as \(\tau_i = Y_i(1) - Y_i(0)\),
where \(Y_i(a)\) represents the potential outcome for unit \(i\) under treatment
condition \(a\) \citep{rubin1974, rubin2005}. Because only one potential outcome can be
observed for each unit, a challenge known as the fundamental problem of causal
inference \citep{holland1986}, causal analyses typically focus on population-level
estimands such as the average treatment effect \citep{rubin2005}. Although randomized
experiments are the gold standard for identifying causal effects \citep{rubin1974},
they are often infeasible, unethical, or impractical in many applied domains,
including medicine, public health, and social sciences. As a result, causal
inference from observational data is ubiquitous. However, causal analyses in
observational settings are fundamentally challenged by confounding bias, in which
covariates jointly influence both treatment assignment and outcomes, leading to
biased effect estimates when not properly addressed \citep{vanderweele2012}.

G-computation, also known as the g-formula \citep{hernan2020}, provides a flexible
approach for estimating causal effects from observational data in the presence
of measured confounding. Originally formalized by Robins, g-computation is
grounded in the potential outcomes framework \citep{robins1986} and proceeds by
modeling the conditional distribution of the outcome given treatment and
observed covariates \citep{hernan2020}. By predicting each individual's potential
outcome under fixed treatment levels and averaging these predictions over the
distribution of observed covariates, g-computation standardizes the predicted
potential outcomes across treatment values. This procedure yields consistent
estimates of causal contrasts, such as differences or ratios of average
potential outcomes, under standard causal identification assumptions, including
consistency, conditional exchangeability, and positivity \citep{bulbulia2024}.

Conventional estimation methods implicitly assume that the observed outcomes of
analysis units are independent. In many applied settings, however, data exhibit
clustered or hierarchical (used interchangeably hereafter) structures, with
individuals nested within higher-level units such as hospitals, schools, or
geographic regions \citep{diez2000, raudenbush2002}. Such nested structures induce
intra-cluster dependence, violating independence assumptions and complicating
causal effect estimation \citep{arpino2011, fanli2013}. Ignoring hierarchical
structure can lead to underestimated standard errors, inflated type I error
rates, and misleadingly precise inference \citep{arceneaux2009}. Random-effects
models (REM), often referred to as hierarchical models, address this dependence
by specifying random effects that capture between-cluster heterogeneity while
pooling information across clusters \citep{diez2000, raudenbush2002}. Consequently,
valid implementation of g-computation in hierarchical settings requires outcome
models that explicitly account for clustering, such as through REM formulations.

This paper aims to formalize the use of g-computation for causal effect
estimation in multilevel data structures and to highlight the necessity of using
REM within g-computation to appropriately account for clustering. However,
hierarchical settings often involve unmeasured cluster-level covariates that
confound the treatment-outcome relationship. In addition, such cluster-level
covariates may induce systematic variation in the treatment effects across
clusters, in which case they act as effect modifiers \citep{vanderweele2012}. Because
these cluster-level variables are unobserved, we can neither directly adjust for
them to fully eliminate confounding bias nor account for treatment effect
heterogeneity across clusters. This challenge is commonly referred to as the
unmeasured context problem \citep{arpino2011}. The primary goal of this paper is to
develop and evaluate a g-computation-based strategy that reduce bias arising
from such unmeasured cluster context in hierarchical observational data.

One intuitive approach to control fully for cluster-level variation, whether
measured or unmeasured, is to apply causal effect estimation methods separately
within each cluster (within-cluster estimation). In many applications, however,
clusters are too small or uneven in size to support stable within-cluster
estimation \citep{lee2021}. To overcome such limitations, Lee et al.~proposed
grouping clusters with similar treatment prevalence and estimating average
treatment effects within each group using group-specific propensity
scores \citep{lee2021}.

Building on this idea, we adopt a within-group estimation strategy in the
context of g-computation to mitigate bias arising from unmeasured cluster-level
confounding and effect modification. Specifically, we propose within-group
g-computation estimators of the treatment effect for both single-level linear
model (LM) and REM. We evaluate their estimation performance through extensive
simulations that vary (i) the number and size of clusters and (ii) key parameters
governing cluster-level confounding and effect modification, comparing these
estimators against standard LM- and REM-based g-computation approaches.

The remainder of the paper is organized as follows. Section \ref{sec-02}
introduces the notation, describes the hierarchical data structure, and states
the identification assumptions required for causal effect estimation from
observational data. Section \ref{sec-03} presents the g-computation framework
under single-level modeling, hierarchical modeling, and within-group estimation.
Section \ref{sec-04} details the simulation design and presents results
evaluating the finite-sample performance of the proposed estimators under
unmeasured cluster-level confounding and treatment effect heterogeneity. Section
\ref{sec-05} applies the proposed within-group REM-based g-computation estimator
to data from the 2019 Bangladesh Multiple Indicator Cluster Survey to estimate
the causal effect of adolescent pregnancy on child linear growth. Section
\ref{sec-06} concludes with a discussion of the findings, methodological
implications, and directions for future research.

\section{Hierarchical Structure and Assumptions}\label{sec-02}

We consider a two-level hierarchical structure in which individuals are nested
within clusters, inducing correlation among outcomes of individuals sharing the
same cluster. We assume that treatment is assigned at the individual level and
the treatment received by one individual does not affect the potential outcomes
of others, either within the same cluster or across clusters.

\begin{figure}[!ht]
\centering
\includegraphics[width=0.3\textwidth]{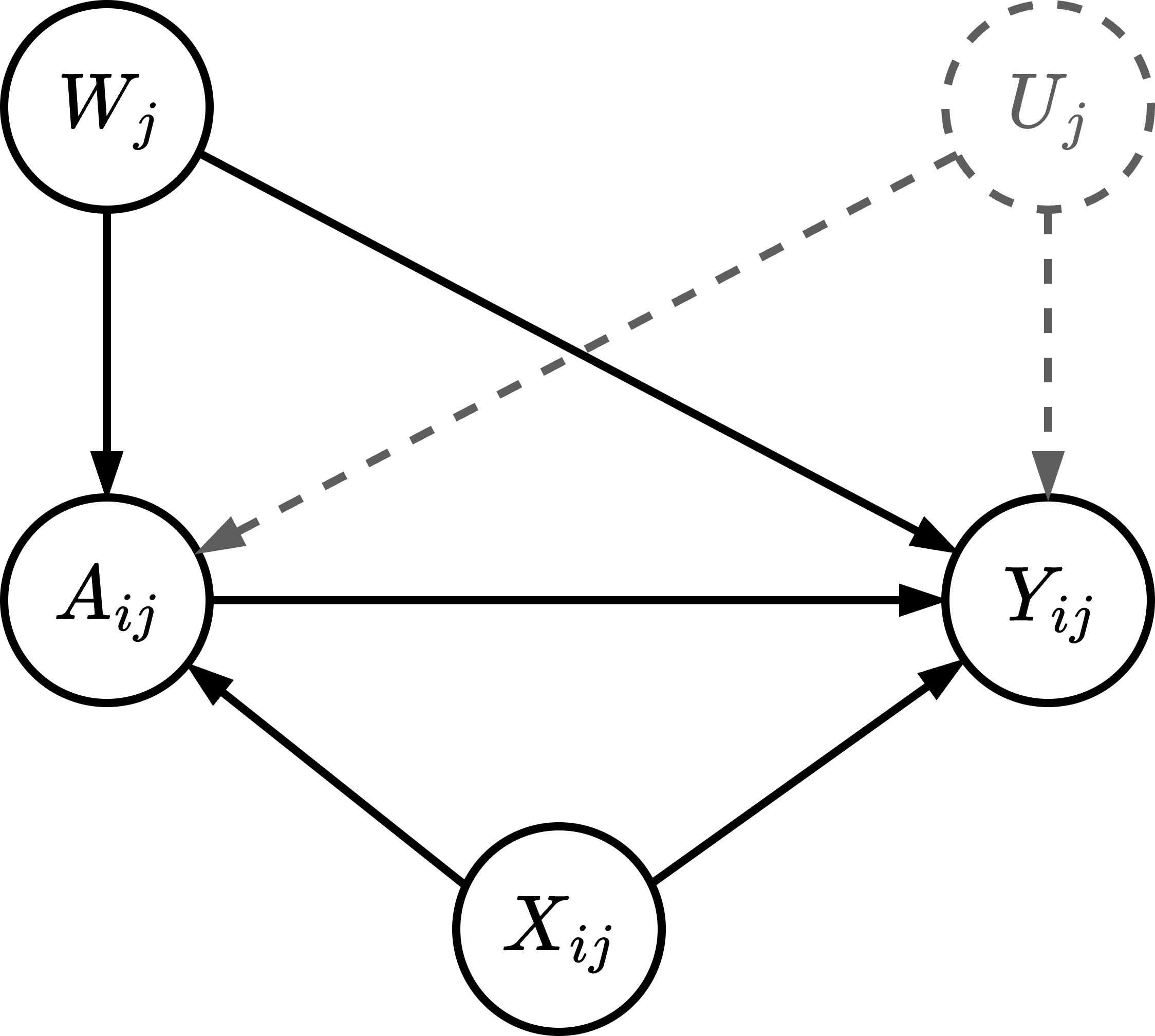}
\caption{Directed acyclic graph illustrating the assumed causal relationships 
among the individual-level covariates $\mathbf{X}_{ij}$, cluster-level 
covariates $\mathbf{W}_j$, treatment $A_{ij}$, and outcome $Y_{ij}$. Solid 
nodes denote observed variables and the dashed node denotes the unmeasured 
cluster-level variable $U_j$, which acts as a confounder by influencing both 
treatment assignment and the outcome.}
\label{fig:dag}
\end{figure}

To formalize this setting, let clusters be indexed by \(j = 1, 2, \dots, J\), where
\(J\) denotes the total number of clusters, and let individuals within cluster \(j\)
be indexed by \(i = 1, \ldots, n_j\), where \(n_j\) is the size of cluster \(j\). For
individual \(i\) in cluster \(j\), let \(A_{ij} \in \{0, 1\}\) denote a binary treatment
(\(A_{ij} = 1\) if treated and \(A_{ij} = 0\) otherwise), \(Y_{ij} \in \mathbb{R}\) the
observed outcome, \(\mathbf{X}_{ij} \in \mathbb{R}^{p}\) individual-level covariates,
and \(\mathbf{W}_j \in \mathbb{R}^{q}\) cluster-level covariates. Both \(\mathbf{X}_{ij}\)
and \(\mathbf{W}_j\) are assumed to be measured prior to treatment assignment. The
observed hierarchical dataset may then be represented as:
\begin{equation*}
\mathcal{D}_{\text{obs}} = \left\{\bigl(Y_{ij}, A_{ij}, \mathbf{X}_{ij}, \mathbf{W}_{j}\bigr) \colon \; i = 1, 2, \dots, n_j; \; j = 1, 2, \dots, J \right\},
\end{equation*}
with total sample size \(n = \sum_{j=1}^{J}n_j\). In addition to the observed
covariates, we posit the existence of an unmeasured cluster-level variable
\(U_j \in \mathbb{R}\), representing contextual characteristics affecting both
\(A_{ij}\) and \(Y_{ij}\) even after adjusting for \(\mathbf{X}_{ij}\) and
\(\mathbf{W}_j\). The assumed causal relationships among these variables are
summarized by the causal directed acyclic graph (DAG) \citep{bulbulia2024} shown in
Figure \ref{fig:dag}.

Let \(\mathcal{L} = \{\mathbf{X}_{ij}, \mathbf{W}_j, U_j\}\) denote the set of
baseline confounders implied by the DAG. The causal estimand of interest is the
average treatment effect (ATE). To define ATE, we use Rubin's potential outcome
framework \citep{rubin2005}. Let \(Y_{ij}(1)\) and \(Y_{ij}(0)\) denote the potential
outcomes of individual \(i\) in cluster \(j\) under treatment and no-treatment,
respectively. We invoke the stable unit-treatment value assumption (SUTVA),
which consists of (i) no interference, meaning that \(Y_{ij}(a)\) for \(a \in \{0, 1\}\)
does not depend on the treatment assignments of other individuals, including
those within the same cluster, and (ii) no multiple versions of treatment
\citep{rubin1980, hernan2020}. Under SUTVA, ATE is defined as the expected difference
between the potential outcomes corresponding to treatment and no-treatment:
\begin{equation*}
\tau = \mathbb{E}\left[Y_{ij}(1) - Y_{ij}(0)\right].
\end{equation*}
Formally, identification of \(\tau\) from the observed data relies on the
following standard causal assumptions \citep{bulbulia2024}:

\begin{enumerate}
  \item Causal consistency: $Y_{ij} = A_{ij}Y_{ij}(1) + (1 - A_{ij})Y_{ij}(0)$
  \item Conditional exchangeability: $\bigl\{Y_{ij}(1), Y_{ij}(0)\bigr\} \perp\!\!\!\perp A_{ij} \mid \mathcal{L}$
  \item Positivity: $0 < \Pr\bigl(A_{ij} = 1 \mid \mathcal{L} \bigr) < 1$
\end{enumerate}

In addition, accurate estimation of \(\tau\) requires that the full set of baseline
confounders \(\mathcal{L}\) be observed in \(\mathcal{D}_{\text{obs}}\)
(i.e.~\(\mathcal{L} \subset \mathcal{D}_{\text{obs}}\)) and that all variables in
\(\mathcal{D}_{\text{obs}}\) are measured without error.

\section{G-Computation in Hierarchical Settings}\label{sec-03}

In this section, we briefly review causal effect estimation using the
g-computation framework and its implementation for hierarchical data. In
particular, we discuss the use of REM within g-computation to account for within-
cluster dependence and highlight the limitation of standard estimation approaches
in the presence of unmeasured cluster-level context, which motivates the within-
group estimation strategy proposed in this paper.

\subsection{The Estimation Framework}\label{the-estimation-framework}

Under the assumptions stated in Section \ref{sec-02}, the mean potential outcome
for treatment assignment \(a \in \{0,1\}\) can be expressed as \citep{hernan2020}:
\begin{equation*}
\mathbb{E}\left[Y_{ij}(a)\right] 
= \mathbb{E}\bigl[\mathbb{E}\left[Y_{ij} \mid A_{ij}=a, \mathcal{L}\right]\bigr]
= \mathbb{E}\left[m(a, \mathcal{L})\right],
\end{equation*}
where \(m(a, \mathcal{L}) = \mathbb{E}\left[Y_{ij} \mid A_{ij} = a, \mathcal{L}\right]\)
denotes the conditional mean outcome. Therefore, our target estimand, ATE, can
be expressed as:
\begin{equation*}
\tau = \mathbb{E}\left[Y_{ij}(1) - Y_{ij}(0)\right] 
= \mathbb{E}\left[m(1, \mathcal{L}) - m(0, \mathcal{L})\right].
\end{equation*}
In practice, estimation of \(\tau\) via g-computation typically proceeds in three
steps. First, an outcome regression model is fit to estimate the conditional
mean \(m(A, \mathcal{L})\). Second, the fitted model is used to predict the
potential outcomes \(\hat{m}(a, \mathcal{L})\) for every individual in the sample
under both treatment conditions \(a \in \{0,1\}\). Finally, the difference in
predicted potential outcomes is averaged to obtain
\(\hat{\tau} = \frac{1}{n} \sum \left\{\hat{m}(1, \mathcal{L}) - \hat{m}(0, \mathcal{L})\right\}\).
Provided that the outcome model \(m(A, \mathcal{L})\) is correctly specified and
all confounders in \(\mathcal{L}\) are observed, \(\hat{\tau}\) is a consistent and
unbiased estimator of \(\tau\) \citep{robins1986, hernan2020}.

\subsection{Outcome Models}\label{outcome-models}

G-computation is a model-agnostic framework and does not require a specific
parametric form for the outcome model. In principle, any regression model may
be used to estimate the conditional mean function \(m(A, \mathcal{L})\). In
single-level settings, this function is commonly modeled using LM. However, when
observations are nested within clusters, outcomes are typically correlated due
to shared cluster-level characteristics, and REM can accommodate this correlation.
Therefore, replacing the single-level LM with an REM as the outcome model extends
g-computation to hierarchical settings by explicitly accounting for within-cluster
correlation.

However, in the setting considered here, the cluster-level variable \(U_j\) is
unobserved, implying that \(\mathcal{L} \not\subset \mathcal{D}_{\text{obs}}\). As
a result, the outcome model is fit using only the observed covariates
\(\{\mathbf{X}_{ij}, \mathbf{W}_j\}\), yielding the estimator
\begin{equation}
\label{eq:full-tau}
\tilde{\tau} 
= 
\frac{1}{n} \sum_{j=1}^{J}\sum_{i=1}^{n_j} 
\left[\hat{m}(1, \mathbf{X}_{ij}, \mathbf{W}_j) 
- 
\hat{m}(0, \mathbf{X}_{ij}, \mathbf{W}_j)\right].
\end{equation}
The assumed role of \(U_j\) in the present setting is consequential, as it defines
the cluster-level context either solely as a confounder or jointly as a
confounder and treatment effect modifier. As we demonstrate in
Section \ref{sec-04}, although REM can substantially reduce bias when \(U_j\)
acts only as a confounder, neither LM nor REM adequately mitigates the bias
when it simultaneously induces treatment effect heterogeneity across clusters.
Consequently, \(\tilde{\tau}\) will generally be a biased estimator of \(\tau\).

This challenge motivates the need for alternative estimation strategies capable
of mitigating bias arising from unmeasured cluster-level context in hierarchical
observational data. In the following subsection, we propose a within-group
g-computation approach designed to address this problem.

\subsection{Within-group Estimation}\label{within-group-estimation}

The within-group g-computation approach proposed in this paper is motivated by
the partially pooled propensity score estimation strategy of Lee et al.~for
inverse probability weighting in hierarchical observational studies \citep{lee2021}.
The rationale for this approach stems from the role of the unmeasured
cluster-level variable \(U_j\): since \(U_j\) influences treatment assignment, the
observed cluster-wise treatment prevalence \(p_j = \sum_{i=1}^{n_j} A_{ij}/n_j\)
encodes information about the latent cluster-level context represented by \(U_j\).
Consequently, clusters with similar treatment prevalence are expected to share
more similar latent context than clusters with substantially different prevalence.
The proposed approach exploits this by grouping clusters with similar \(p_j\) and
performing g-computation within each group, thereby reducing between-cluster
variation in \(U_j\) within groups.

A variety of grouping methods can be used to form groups of clusters based on
\(p_j\). In this study, we employ the k-means algorithm \citep{hartigan1979},
which partitions the \(J\) clusters into \(G < J\) mutually exclusive groups. Let
\(g(j) \in \{1,\dots,G\}\) denote the group membership of cluster \(j\). After
forming the groups, g-computation is performed within each group. Specifically,
for each group \(g\), an outcome model \(m_g(A_{ij}, \mathbf{X}_{ij}, \mathbf{W}_j)\)
is fitted using only observations belonging to clusters assigned to that group.
The fitted model is then used to predict the potential outcomes under both
treatment conditions for all individuals within the same group, yielding the
group-specific estimator:
\begin{equation}
\label{eq:grp-tau}
\tilde{\tau}_g
=
\frac{1}{n_g}
\sum_{j:g(j)=g}\sum_{i=1}^{n_j}
\left[
\hat{m}_g(1,\mathbf{X}_{ij},\mathbf{W}_j)
-
\hat{m}_g(0,\mathbf{X}_{ij},\mathbf{W}_j)
\right],
\end{equation}
where \(n_g = \sum_{j:g(j)=g} n_j\) denotes the total number of individuals in
group \(g\). Finally, the marginal ATE is estimated as the weighted average of the
group-specific estimators:
\begin{equation}
\label{eq:wg-tau}
\tilde{\tau}_{WG} = \sum_{g=1}^{G} w_g\tilde{\tau}_g,
\end{equation}
where \(w_g = n_g/n\) corresponds to the proportion of individuals
belonging to group \(g\). This weighting scheme ensures that each group contributes
to the marginal estimator in proportion to its size, thereby preserving the
interpretation of \(\tilde{\tau}_{WG}\) as an estimator of the population ATE.

This within-group estimation strategy can be implemented using either LM or REM
for the group-specific outcome models. However, for hierarchical data, the
REM-based within-group estimator is expected to be less susceptible to bias
arising from unmeasured cluster-level context than the corresponding LM-based
within-group estimator. This is because, while the grouping strategy is intended
to reduce variation in latent cluster-level factors within each group, REM
can additionally account for the within-cluster outcome correlation.

\section{Simulation Study}\label{sec-04}

We conducted a Monte Carlo (MC) simulation study to evaluate the finite-sample
performance of the proposed within-group g-computation estimators in a two-level
hierarchical structure described in Section \ref{sec-02}. Two primary simulation
scenarios were considered, corresponding to different roles of the unmeasured
cluster-level variable \(U_j\): (i) acting solely as a confounder and (ii) acting
jointly as a confounder and a treatment effect modifier. Across all scenarios,
data were generated in accordance with the causal structure depicted in
Figure \ref{fig:dag}.

For individuals \(i = 1, \dots, n_j\) nested within clusters \(j = 1, \dots, J\),
the covariates \(X_{ij}\), \(W_j\), and \(U_j\) were independently generated from
\(\mathcal{N}(0,1)\). Treatment assignment \(A_{ij}\) was drawn from a Bernoulli
distribution with probability \(\pi_{ij} = \Pr(A_{ij} = 1 \mid \mathcal{L})\)
defined as:
\begin{equation*}
\text{logit}(\pi_{ij}) = \gamma_{0j} + \gamma_1 X_{ij} + \gamma_2 W_j + \gamma_3 U_j,
\end{equation*}
where \(\gamma_1 = \gamma_2 = -1\) and \(\gamma_{0j} \sim \mathcal{N}(0, 0.25)\).
The parameter \(\gamma_3\), which controls the influence of \(U_j\) on treatment
assignment, was varied across simulation settings. The potential outcomes under
no-treatment and treatment conditions were defined as:
\begin{equation*}
\begin{aligned}
Y_{ij}(0) &= \beta_{0j} + \beta_1 X_{ij} + \beta_2 W_j + \beta_3 U_j + \varepsilon_{ij}, \\
Y_{ij}(1) &= Y_{ij}(0) + \zeta + \theta U_j^2,
\end{aligned}
\end{equation*}
where \(\beta_1 = \beta_2 = 1\), \(\beta_{0j} \sim \mathcal{N}(5,1)\), and
\(\varepsilon_{ij} \sim \mathcal{N}(0,1)\). The parameter \(\beta_3\), governing the
influence of \(U_j\) on the outcome, was also varied across simulation settings.
The resulting individual treatment effect is:
\begin{equation*}
Y_{ij}(1) - Y_{ij}(0) = \zeta + \theta U_j^2.
\end{equation*}
To ensure a constant marginal ATE of \(\tau=1\) across all simulation scenarios,
we defined \(\zeta = 1 - \theta \mathbb{E}(U_j^2)\). Observed outcomes were
generated via the consistency relation:
\begin{equation*}
\begin{aligned}
Y_{ij} &= A_{ij}Y_{ij}(1) + (1 - A_{ij})Y_{ij}(0) \\
       &= Y_{ij}(0) + A_{ij} \left(\zeta + \theta U_j^2\right).
\end{aligned}
\end{equation*}
The parameter \(\theta\) determines the degree of treatment effect heterogeneity
induced by \(U_j\), whereas the parameter set \((\gamma_3, \beta_3)\) controls the
strength of unmeasured cluster-level confounding. To ensure adequate covariate
overlap, we retained only clusters satisfying \(p_j \in (0.05, 0.95)\).

Across all simulation settings, we compared four estimators: the standard LM
estimator \(\tilde{\tau}_{\text{LM}}\), the standard REM estimator
\(\tilde{\tau}_{\text{REM}}\), and their within-group counterparts
\(\tilde{\tau}_{\text{WG-LM}}\) and \(\tilde{\tau}_{\text{WG-REM}}\). Estimates for
the standard estimators were obtained using Equation \eqref{eq:full-tau} after
fitting the outcome model to the entire dataset. For within-group estimation,
clusters were partitioned into \(G = 5\) groups using k-means clustering on \(p_j\)
and the outcome model was fitted separately within each group. Group-specific
ATE estimates were obtained using Equation \eqref{eq:grp-tau} and combined via
Equation \eqref{eq:wg-tau} to yield the final within-group estimator. In all
cases, the outcome model was specified as:
\begin{equation*}
\begin{aligned}
\text{LM}: \; Y_{ij} &= 
\alpha_0 + \alpha_1 A_{ij} + \alpha_2 X_{ij} + \alpha_3 W_j + \epsilon_{ij}, \\
\text{REM}: \; Y_{ij} &= 
\alpha_{0} + \alpha_1 A_{ij} + \alpha_2 X_{ij} + \alpha_3 W_j + b_j + \epsilon_{ij},
\end{aligned}
\end{equation*}
where \(b_j \sim \mathcal{N}(0, \sigma_b^2)\) and
\(\epsilon_{ij} \sim \mathcal{N}(0, \sigma_\epsilon^2)\).

We replicated each simulation setting 500 times and assessed the performance of
each estimator using bias, standard deviation (SD), and root mean squared error
(RMSE), defined as:
\begin{equation*}
\begin{aligned}
\text{Bias}  &= \frac{1}{500}\sum_{r=1}^{500}\left(\tilde{\tau}^{(r)} - \tau\right), \\
\text{SD}    &= \sqrt{\frac{1}{499}\sum_{r=1}^{500}\left(\tilde{\tau}^{(r)} - 
                \bar{\tilde{\tau}}\right)^2}, \\[2pt]
\text{RMSE}  &= \sqrt{\frac{1}{500}\sum_{r=1}^{500}\left(\tilde{\tau}^{(r)} - \tau\right)^2},
\end{aligned}
\end{equation*}
where \(\tilde{\tau}^{(r)}\) denotes the estimated ATE from the \(r\)th MC
replication and \(\bar{\tilde{\tau}}\) is its mean across all replications.
Since RMSE captures both bias and variability, the simulation results are
presented using RMSE in this article, with the corresponding bias and SD
provided in the Supplementary Material
(Tables \ref{tab:s1-gb}, \ref{tab:s1-cls}, \ref{tab:s2-theta}, and \ref{tab:s2-cls}).

\subsection{Scenario 1: Unmeasured Cluster-level Confounding}\label{scenario-1-unmeasured-cluster-level-confounding}

In this scenario, we set \(\theta = 0\), imposing a constant treatment effect
across clusters. Consequently, the unmeasured cluster-level variable \(U_j\)
acted solely as a confounder rather than a treatment effect modifier.

\begin{figure}[!ht]
\centering
\includegraphics[width=0.80\textwidth]{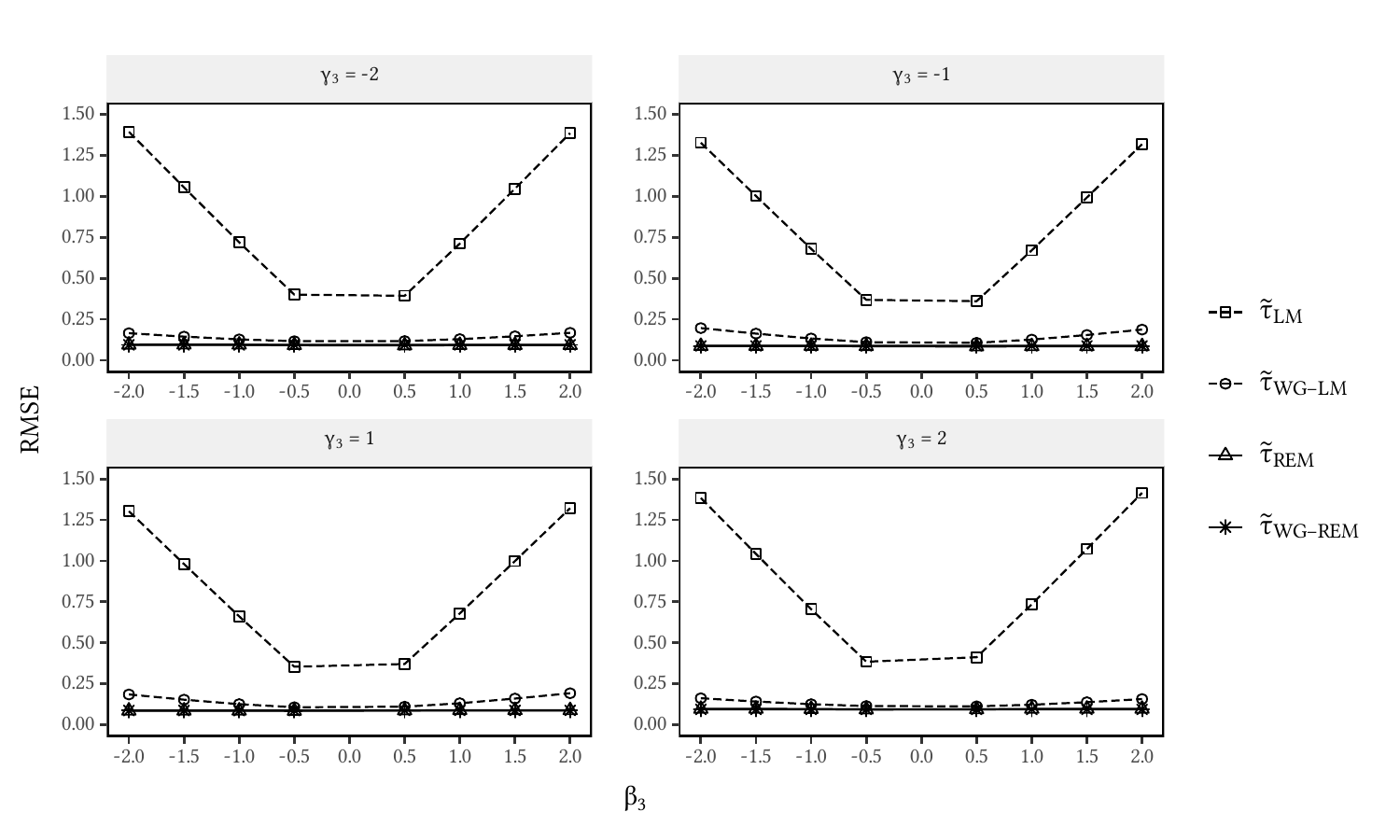}
\caption{Root mean squared error (RMSE) of the four g-computation estimators
as a function of the outcome confounding parameter
$\beta_3 \in \{-2, -1.5, -1, -0.5, 0.5, 1, 1.5, 2\}$, across four values of
the treatment confounding parameter $\gamma_3 \in \{-2, -1, 1, 2\}$ (panels).
$\tilde{\tau}_{\text{LM}}$: standard estimator based on linear model; 
$\tilde{\tau}_{\text{REM}}$: standard estimator based on random-effects model; 
$\tilde{\tau}_{\text{WG-LM}}$: within-group estimator based on linear model;
$\tilde{\tau}_{\text{WG-REM}}$: within-group estimator based on random-effects 
model. Results correspond to Scenario 1 ($\theta = 0$), with $J = 50$ clusters 
of size $n_j = 20$. For within-group estimators, clusters were partitioned into 
$G = 5$ groups based on cluster-wise treatment prevalence. The true average 
treatment effect is $\tau = 1$.}
\label{fig:s-1-gb}
\end{figure}

To assess how estimator performance varies with the magnitude of confounding, we
fixed \(J = 50\), \(n_j = 20\) and measured RMSE across
\(\beta_3 \in \{-2,-1.5,-1,-0.5,0.5,1,1.5,2\}\) for each \(\gamma_3 \in \{-2,-1,1,2\}\).
Figure \ref{fig:s-1-gb} presents the RMSE of each estimator as a function of
\(\beta_3\), with each panel corresponding to a fixed value of \(\gamma_3\). As
\(\lvert\beta_3\rvert\) increased, \(\tilde{\tau}_{\text{LM}}\) showed increasing
RMSE across all panels, whereas \(\tilde{\tau}_{\text{WG-LM}}\) exhibited a similar
but markedly attenuated pattern, indicating notable mitigation of confounding
bias. In contrast, both REM-based estimators maintained uniformly low RMSE
across all values of \(\gamma_3\) and \(\beta_3\). Their nearly identical performance
suggests that, under homogeneous treatment effects, REM can effectively account
for confounding induced by \(U_j\), rendering within-group estimation largely
redundant in this setting.

\begin{figure}[!ht]
\centering
\includegraphics[width=0.80\textwidth]{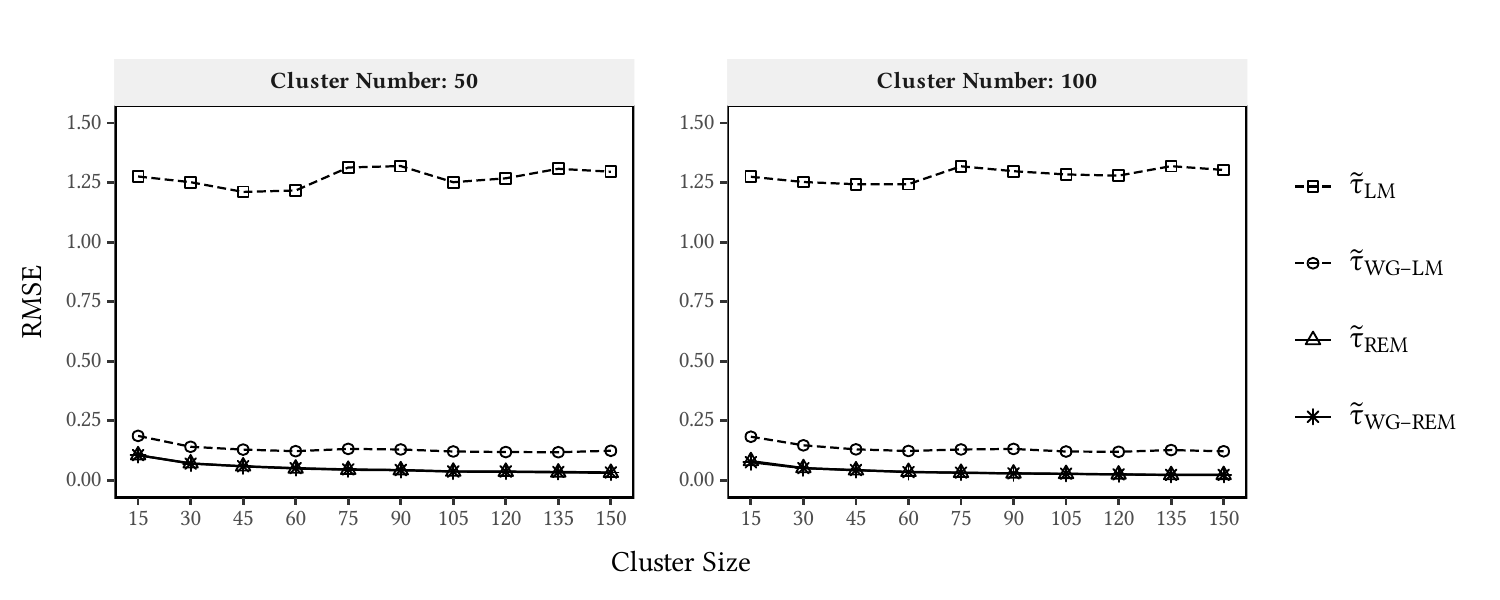}
\caption{Root mean squared error (RMSE) of the four g-computation estimators
across cluster sizes $n_j \in \{15, 30, \ldots, 150\}$, for $J = 50$ clusters
(left panel) and $J = 100$ clusters (right panel). 
$\tilde{\tau}_{\text{LM}}$: standard estimator based on linear model; 
$\tilde{\tau}_{\text{REM}}$: standard estimator based on random-effects model; 
$\tilde{\tau}_{\text{WG-LM}}$: within-group estimator based on linear model;
$\tilde{\tau}_{\text{WG-REM}}$: within-group estimator based on random-effects 
model. For within-group estimators, clusters were partitioned into $G = 5$ 
groups based on cluster-wise treatment prevalence. Results correspond to 
Scenario 1 ($\theta = 0$), with confounding parameters fixed at 
$\gamma_3 = \beta_3 = 1.5$. The true average treatment effect is $\tau = 1$.}
\label{fig:s-1-cls}
\end{figure}

To examine whether these patterns persisted across varying numbers of clusters
and cluster sizes, we fixed \(\gamma_3 = \beta_3 = 1.5\) and varied \(n_j\) over
\(\{15, 30, \ldots, 150\}\) for \(J \in \{50, 100\}\). Figure \ref{fig:s-1-cls}
confirms the same ordering of estimators across all combinations of \(n_j\) and
\(J\): \(\tilde{\tau}_{\text{LM}}\) remained the worst performer, whereas
\(\tilde{\tau}_{\text{WG-LM}}\) offered nontrivial improvement. Both REM-based
estimators achieved considerably lower RMSE with nearly identical performance,
with RMSE declining steadily as cluster size increased. Increasing the number of
clusters had minimal impact on the relative performance.

\subsection{Scenario 2: Unmeasured Cluster-level Confounding and Effect Modification}\label{scenario-2-unmeasured-cluster-level-confounding-and-effect-modification}

\begin{figure}[!ht]
\centering
\includegraphics[width=0.75\textwidth]{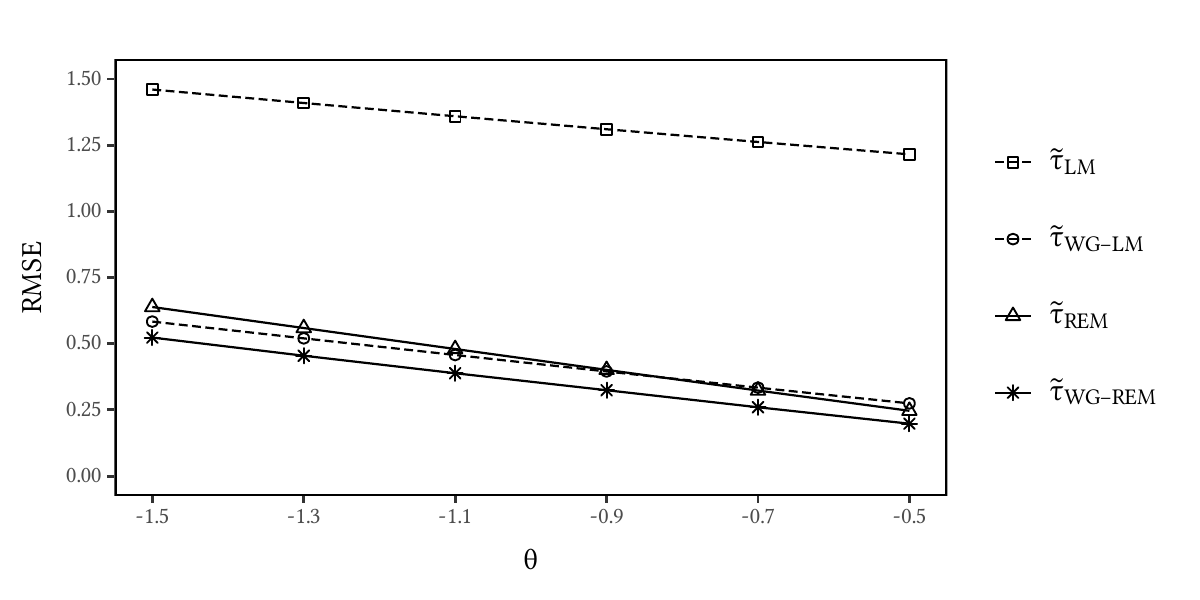}
\caption{Root mean squared error (RMSE) of the four g-computation estimators
as a function of the effect heterogeneity parameter
$\theta \in \{-1.5, -1, -0.5, 0.5, 1, 1.5\}$. 
$\tilde{\tau}_{\text{LM}}$: standard estimator based on linear model; 
$\tilde{\tau}_{\text{REM}}$: standard estimator based on random-effects model; 
$\tilde{\tau}_{\text{WG-LM}}$: within-group estimator based on linear model;
$\tilde{\tau}_{\text{WG-REM}}$: within-group estimator based on random-effects 
model. Results correspond to Scenario 2, with $J = 50$ clusters of size $n_j = 20$ 
and confounding parameters fixed at $\gamma_3 = \beta_3 = 1.5$. For within-group 
estimators, clusters were partitioned into $G = 5$ groups based on cluster-wise 
treatment prevalence. The true average treatment effect is $\tau = 1$.}
\label{fig:s-2-theta}
\end{figure}

In this scenario, we set \(\theta \neq 0\), so that \(U_j\) simultaneously acted
as a confounder and a treatment effect modifier. To assess how estimator
performance varies with the degree of treatment effect heterogeneity, we fixed
\(J = 50\), \(n_j = 20\), \(\gamma_3 = \beta_3 = 1.5\), and varied
\(\theta \in \{-1.5, -1, -0.5, 0.5, 1, 1.5\}\). Figure \ref{fig:s-2-theta}
presents the RMSE of each estimator as a function of \(\theta\). As
\(\lvert\theta\rvert\) increased, all estimators showed higher RMSE, reflecting
the increasing treatment effect heterogeneity induced by \(U_j\). However, the
pattern of relative performance differed markedly from Scenario 1. Both
\(\tilde{\tau}_{\text{WG-LM}}\) and \(\tilde{\tau}_{\text{WG-REM}}\) achieved
considerably lower RMSE than their standard counterparts
\(\tilde{\tau}_{\text{LM}}\) and \(\tilde{\tau}_{\text{REM}}\), demonstrating that
within-group estimation offers meaningful bias reduction when \(U_j\) induces
treatment effect heterogeneity. Notably, \(\tilde{\tau}_{\text{WG-REM}}\)
consistently achieved the lowest RMSE across all values of \(\theta\), suggesting
that combining within-group estimation with random-effects adjustment provides
the most effective mitigation of bias under unmeasured cluster-level context.

\begin{figure}[!ht]
\centering
\includegraphics[width=0.85\textwidth]{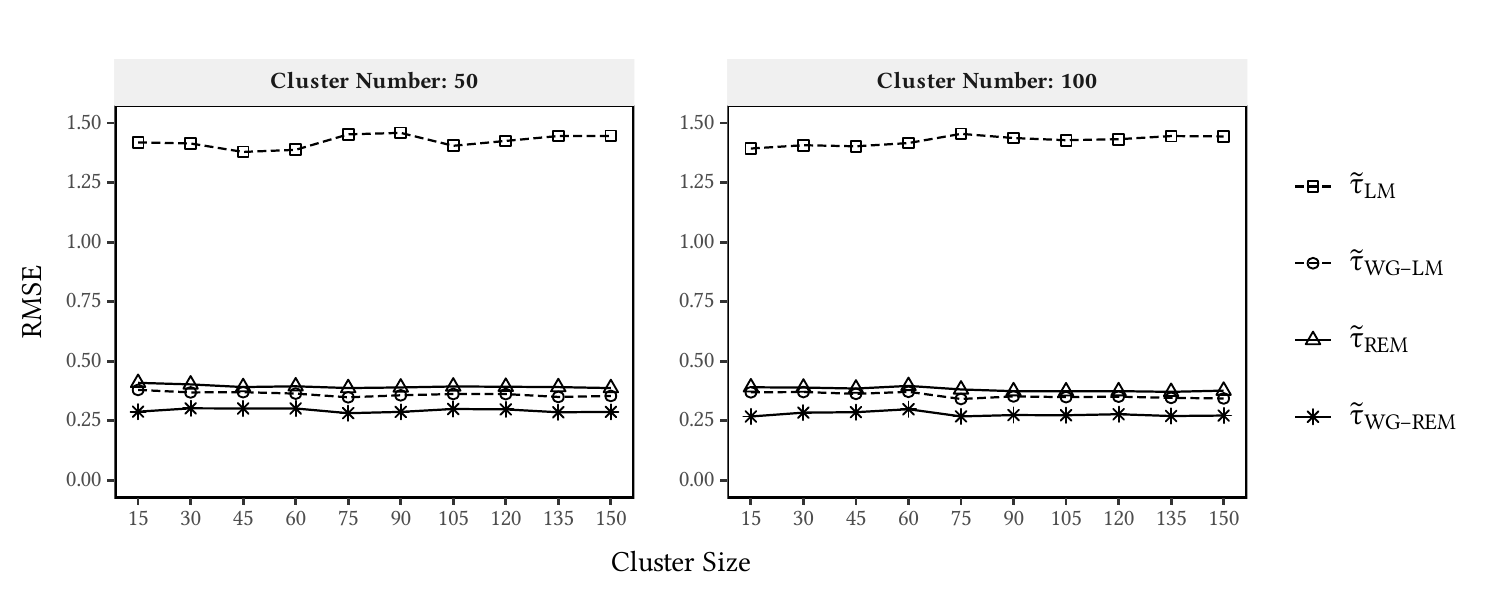}
\caption{Root mean squared error (RMSE) of the four g-computation estimators
across cluster sizes $n_j \in \{15, 30, \ldots, 150\}$, for $J = 50$ clusters
(left panel) and $J = 100$ clusters (right panel). 
$\tilde{\tau}_{\text{LM}}$: standard estimator based on linear model; 
$\tilde{\tau}_{\text{REM}}$: standard estimator based on random-effects model; 
$\tilde{\tau}_{\text{WG-LM}}$: within-group estimator based on linear model;
$\tilde{\tau}_{\text{WG-REM}}$: within-group estimator based on random-effects 
model. Results correspond to Scenario 2 ($\theta = -1$), with confounding 
parameters fixed at $\gamma_3 = \beta_3 = 1.5$. For within-group estimators, 
clusters were partitioned into $G = 5$ groups based on cluster-wise treatment 
prevalence. The true average treatment effect is $\tau = 1$.}
\label{fig:s-2-cls}
\end{figure}

To examine whether these patterns persisted across varying cluster sizes and
numbers of clusters, we fixed \(\gamma_3 = \beta_3 = 1.5\), \(\theta = -1\), and
varied \(n_j\) over \(\{15, 30, \ldots, 150\}\) for \(J \in \{50, 100\}\).
Figure \ref{fig:s-2-cls} confirms the same ordering of estimators across all
combinations of \(n_j\) and \(J\): the within-group estimators consistently
outperformed their standard counterparts, with \(\tilde{\tau}_{\text{WG-REM}}\)
achieving the lowest RMSE overall, and RMSE declining steadily with cluster size.

\section{Real Data Application}\label{sec-05}

To illustrate the proposed within-group g-computation framework in an applied
setting, we investigated the causal effect of adolescent pregnancy (ADP) on
child linear growth in Bangladesh. Child undernutrition remains a major public
health concern and a central target of global development policy. Sustainable
Development Goal (SDG) 2.2 called for a 40\% reduction in the prevalence of
stunting among children under five by 2025 \citep{mics2019}, recognizing that
improvements in early childhood nutrition are essential for child survival,
cognitive development, and long-term human capital formation \citep{li2023}.

Nutritional status among children under five is commonly assessed using
standardized anthropometric indicators based on the World Health Organization
(WHO) growth standards, including weight-for-age, weight-for-height, and
height-for-age Z-scores \citep{who2006}. We focused on height-for-age Z-score (HAZ),
which measures linear growth relative to a healthy reference population and is
widely used to assess chronic undernutrition in early childhood \citep{li2023}.
Children with HAZ below -2 standard deviations from the WHO reference median
are classified as stunted, indicating moderate to severe chronic undernutrition
with well-documented adverse consequences for cognitive development, educational
attainment, and long-term health \citep{li2023}.

ADP, defined as first childbirth occurring at age 19 years or younger
\citep{islam2017, alam2018}, has been identified in the literature as an important
determinant of impaired child growth and stunting \citep{nguyen2019, azriani2024, welch2024}.
Biologically, adolescent mothers may still be undergoing physical growth
themselves, potentially creating competition for nutritional resources between
the mother and fetus \citep{wallace2019}. In addition, ADP is strongly associated
with socioeconomic disadvantage, limited healthcare access, and poorer maternal
health behaviors, all of which may adversely affect child nutritional
outcomes \citep{welch2024}.

\subsection{Data Source and Variables}\label{data-source-and-variables}

The analysis used secondary data from the 2019 Bangladesh Multiple Indicator
Cluster Survey (MICS), which is a nationally representative cross-sectional
household survey conducted by the Bangladesh Bureau of Statistics in
collaboration with UNICEF as part of the global MICS programme. Data collection
was carried out between January and June 2019 using standardized questionnaires
administered to women aged 15-49 years and caregivers of children under five,
capturing information on maternal and child health, nutrition, education, and
household socioeconomic conditions \citep{mics2019}. The survey employed a two-stage
stratified cluster sampling design, covering 64,400 households nationwide. Full
details of the sampling design and recorded variables are provided in the
official survey report \citep{mics2019}. The survey data were anonymized prior to
public release to protect respondent confidentiality and are freely available
through the MICS website (\url{https://mics.unicef.org/surveys}).

The \(J = 64\) administrative districts of Bangladesh were considered as clusters,
with mother-child pairs constituting the individual-level units nested within
districts. The exposure variable was constructed as
\(\text{ADP} = \mathbb{I}(\text{age at first birth} \leq 19)\), where the mother's
age at first birth was computed as the difference between her reported birth
year and the year of her first delivery, both recorded in the survey. The
outcome variable (HAZ) was directly available from the survey data. Following
WHO recommendations, biologically implausible HAZ values outside the interval
(-6, 6) were excluded from the analysis \citep{who2019}.

To avoid within-family dependence arising from multiple children per mother,
only the oldest child from each mother was retained, ensuring one observation
per mother. Since the oldest child is necessarily the firstborn, this selection
also ensured that the observed HAZ corresponded to the child whose birth defined
the exposure, thereby preserving the temporal ordering between exposure and
outcome despite the cross-sectional nature of the survey. The final analytic
sample consisted of 19,424 mother-child pairs with complete information on the
outcome, exposure, and all measured covariates. Three individual-level baseline
covariates associated with both ADP and child HAZ were included as measured
confounders: maternal education level (binary: secondary or higher vs.~less than
secondary), household wealth index (categorical: poor, middle, rich), and place
of residence (binary: urban vs.~
rural) \citep{islam2017, kassa2018, alam2018, chowdhury2022, kundu2025}.

In Bangladesh, maternal and child health services vary across districts due to
differences in healthcare infrastructure, availability of healthcare
professionals and community health workers, and patterns of service utilization.
These district-level conditions have been shown to influence both child stunting
\citep{azriani2024, astuti2025} and ADP \citep{nguyen2019, vieira2023}. However,
information on healthcare availability and utilization was not directly
observed in the survey data. Such unmeasured district-level contextual factors
may constitute a plausible source of cluster-level confounding and may
additionally induce treatment effect heterogeneity across districts.
As the simulation study in Section \ref{sec-04} showed that the within-group
REM-based g-computation estimator achieved the lowest RMSE under such conditions,
we used this estimator in the current analysis.

\subsection{Estimation and Results}\label{estimation-and-results}

\begin{figure}[!ht]
\centering
\includegraphics[width=0.5\textwidth]{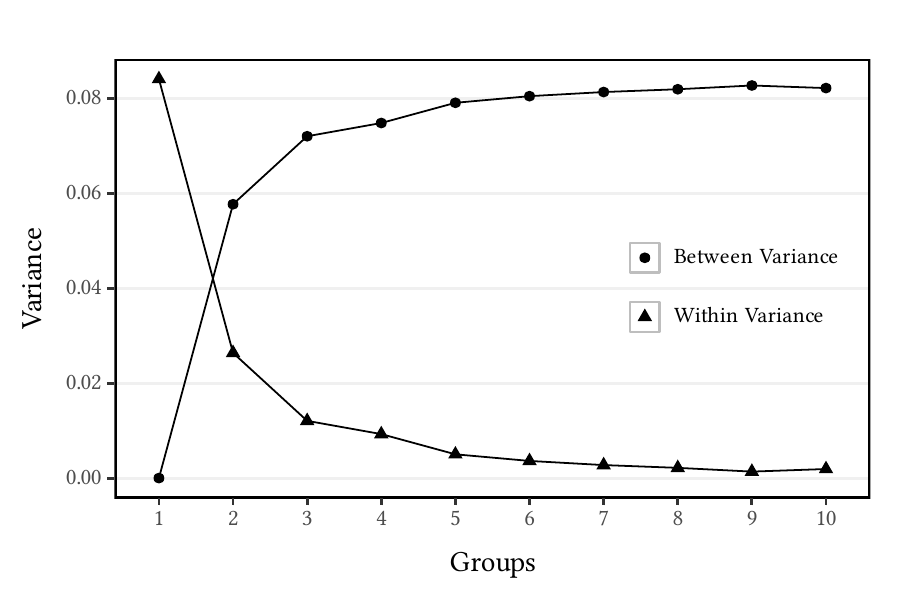}
\caption{Mean prevalence of ADP (adolescent pregnancy) across groups for 
different values of $G$ (number of groups). Each panel corresponds to a 
specific value of $G$ as indicated by the panel label. The $x$-axis denotes 
the group ID assigned by the k-means algorithm and numbers above each bar 
indicate the number of districts within that group.}
\label{fig:grps-bar}
\end{figure}

\begin{figure}[!ht]
\centering
\includegraphics[width=0.9\textwidth]{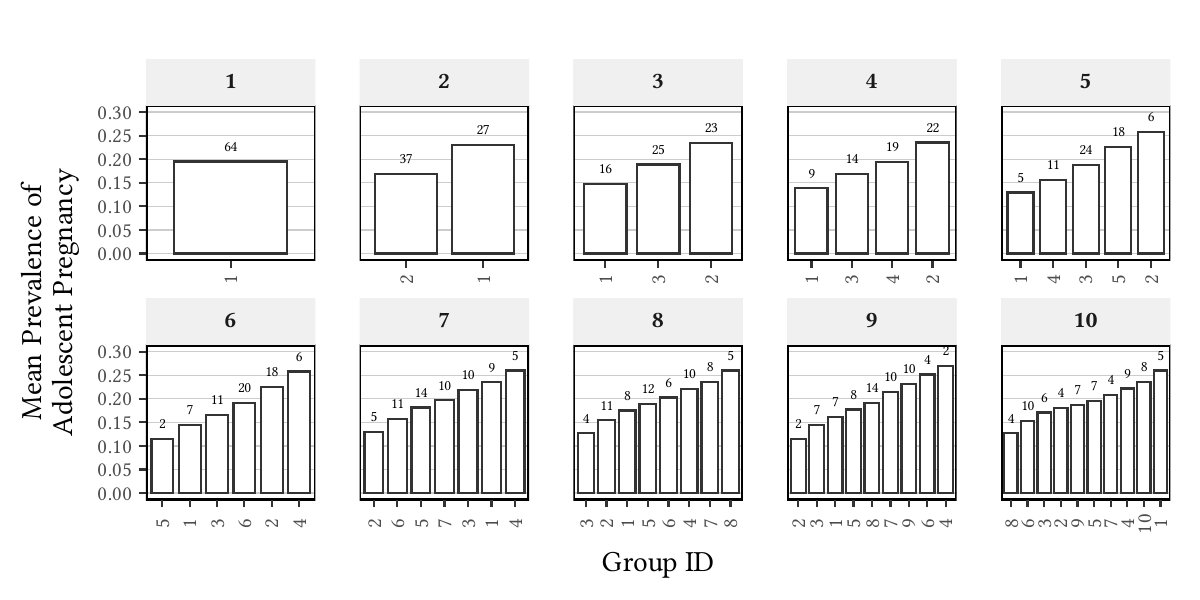}
\caption{Between- and within-group variance of district-level ADP prevalence 
across different values of $G$ (number of groups). Filled circles represent 
between-group variance and filled triangles represent within-group variance.}
\label{fig:elbow-plot}
\end{figure}

Following the methodology in Section \ref{sec-03}, districts were grouped
according to their observed ADP prevalence using the k-means algorithm. Figures
\ref{fig:grps-bar} and \ref{fig:elbow-plot} provide a visual assessment of the
within-group homogeneity and between-group heterogeneity in ADP prevalence under
different values of \(G\) and informed the selection of \(G = 5\) groups, balancing
within-group homogeneity in ADP prevalence against a sufficient number of
districts per group.

Within each group, a weighted REM with a district-specific random intercept
was fitted, adjusting for the three measured confounders and incorporating the
survey sampling weights to account for the complex survey design \citep{rabe2006}.
The marginal ATE was then obtained by aggregating the group-specific estimates
according to Equation \eqref{eq:wg-tau}. A 95\% confidence interval (CI) was
constructed from the 2.5th and 97.5th percentiles of 1,000 bootstrap estimates.
Bootstrap resampling was performed at the level of the survey primary sampling
unit (enumeration area), distinct from the analytical clusters (districts), to
respect the complex survey design. The distribution of the bootstrap estimates
is displayed in Supplementary Figure \ref{fig:boot-hist-plot}.

The within-group REM based g-computation estimator yielded an ATE of -0.12
(95\% bootstrap CI: {[}-0.18, -0.06{]}), indicating that children born to
adolescent mothers had HAZ approximately 0.12 standard deviation units
lower, on average, than children born to mothers whose first pregnancy
occurred at age 20 or older, after adjusting for measured confounders. The
confidence interval excluded zero, providing evidence against a null effect.
The estimated effect is consistent with the epidemiological literature linking
ADP to poorer child growth outcomes \citep{nguyen2019, azriani2024, welch2024}, and
suggests that ADP has meaningful long-term implications for child nutritional
health in Bangladesh.

\section{Conclusion}\label{sec-06}

This paper examined causal effect estimation from observational hierarchical
data using the g-computation framework. We first discussed the use of REM as
the outcome model in g-computation to account for within-cluster dependence,
extending the LM-based implementation to hierarchical settings. We then proposed
a within-group g-computation framework to address settings in which unmeasured
cluster-level factors influence both treatment assignment and outcomes. The
proposed approach combines cluster grouping by treatment prevalence with
group-specific outcome modeling using either LM or REM, leading to within-group
estimators designed to reduce bias arising from unmeasured cluster-level
confounding and treatment effect heterogeneity across clusters.

Through extensive Monte Carlo simulations, we evaluated four estimators:
standard LM-based g-computation, standard REM-based g-computation, and their
within-group counterparts. When the unmeasured cluster-level variable acted
solely as a confounder, both REM-based estimators exhibited similarly low bias
and RMSE, indicating that random-effects adjustment can adequately account for
latent cluster-level confounder under homogeneous treatment effects. In contrast,
when the unmeasured confounder also induced treatment effect heterogeneity, the
within-group REM-based estimator consistently achieved the lowest RMSE among
all competing approaches. This advantage was observed across varying numbers and
sizes of clusters.

To illustrate the practical utility of the proposed approach, the within-group
REM-based estimator was applied to data from the 2019 Bangladesh MICS to estimate
the causal effect of adolescent pregnancy on child linear growth. Under the
identifying assumptions, adolescent pregnancy was estimated to reduce child HAZ
by approximately 0.12 standard deviation units on average (95\% bootstrap CI:
{[}-0.18, -0.06{]}), after adjustment for measured confounders. This finding is
consistent with previous epidemiological evidence linking adolescent pregnancy
to adverse child growth outcomes and highlights the practical relevance of the
proposed framework for addressing public health questions in which unmeasured
district-level contextual factors are likely to confound causal estimates.

Several limitations remain. First, the simulation study focused on linear
outcome-generating mechanisms, and the performance of the proposed estimator
under nonlinear relationships, non-Gaussian outcomes, or more complex
hierarchical structures remains to be evaluated. Second, the number of groups
was fixed at five in both the simulation study and the real data application,
with the latter choice informed by a visual heuristic based on within- and
between-group variance. The optimal number of groups may vary across settings,
and future research should investigate data-driven dynamic procedures for
selecting this tuning parameter and evaluate the sensitivity of the estimator
to its choice. Finally, although k-means clustering was used to group clusters
by treatment prevalence, alternative grouping strategies may offer improved
performance in certain settings and require further investigation.

In summary, the proposed framework suggests that treatment prevalence at the
cluster level may contain information about unmeasured contextual factors,
and that exploiting this information through within-group estimation can
meaningfully improve causal effect estimation in hierarchical observational
studies with latent cluster context. As observational data from clustered or
multilevel designs become increasingly prevalent in epidemiology, public health,
and the social sciences, methods that remain robust to unmeasured cluster
context are of growing importance. The proposed within-group g-computation
framework provides a simple and effective approach for such settings.

\section*{Acknowledgements}\label{acknowledgements}
\addcontentsline{toc}{section}{Acknowledgements}

\vspace{-1em}

The authors thank the MICS programme for providing free access to the survey data.

\section*{Author's Contributions}\label{authors-contributions}
\addcontentsline{toc}{section}{Author's Contributions}

\vspace{-1em}

S.K.S.: Conceptualization, Data curation, Methodology, Formal analysis, Software,
Visualization, Writing - original draft. B.S.: Methodology, Writing - original
draft. M.N.I.S.: Conceptualization, Methodology, Writing - original draft. All
authors reviewed and approved the final manuscript.

\section*{Funding}\label{funding}
\addcontentsline{toc}{section}{Funding}

\vspace{-1em}

The authors received no specific funding for this work.

\section*{Competing Interests}\label{competing-interests}
\addcontentsline{toc}{section}{Competing Interests}

\vspace{-1em}

The authors declare no competing interests.

\section*{Data Availability}\label{data-availability}
\addcontentsline{toc}{section}{Data Availability}

\vspace{-1em}

The data used in the real data application are freely available through the
MICS website at \url{https://mics.unicef.org/surveys}.

\section*{ORCID}\label{orcid}
\addcontentsline{toc}{section}{ORCID}

\vspace{-1em}

Shafayet Khan Shafee \orcidlink{0009-0002-8021-3788} \url{https://orcid.org/0009-0002-8021-3788}

Bishal Sarker \orcidlink{0009-0001-0812-4142} \url{https://orcid.org/0009-0001-0812-4142}

Md. Niamul Islam Sium \orcidlink{0009-0005-1822-8294} \url{https://orcid.org/0009-0005-1822-8294}

\renewcommand\refname{References}

\newpage
\section*{Supplementary Materials}

\setcounter{table}{0}
\renewcommand{\thetable}{S\arabic{table}}
\setcounter{figure}{0}
\renewcommand{\thefigure}{S\arabic{figure}}

\captionsetup[table]{skip=10pt}

\begin{table}[!h]
\centering
\caption{\label{tab:s1-gb}Simulation results for Scenario 1 (unmeasured cluster-level confounding only with $\theta = 0$): bias and standard deviation (SD) of the four g-computation estimators across combinations of $\gamma_3$ and $\beta_3$, with $J = 50$ clusters and $n_j = 20$ individuals per cluster based on 500 Monte Carlo replications. $\tilde{\tau}_{\text{LM}}$, $\tilde{\tau}_{\text{WG-LM}}$, $\tilde{\tau}_{\text{REM}}$, and $\tilde{\tau}_{\text{WG-REM}}$ denote the standard linear model, within-group linear model, standard random-effects model, and within-group random-effects g-computation estimators, respectively.}
\centering
\fontsize{9}{11}\selectfont
\begin{threeparttable}
\begin{tabular}[t]{rrrrrrrrrr}
\toprule
\multicolumn{2}{c}{ } & \multicolumn{2}{c}{$\tilde{\tau}_{\text{LM}}$} & \multicolumn{2}{c}{$\tilde{\tau}_{\text{WG-LM}}$} & \multicolumn{2}{c}{$\tilde{\tau}_{\text{REM}}$} & \multicolumn{2}{c}{$\tilde{\tau}_{\text{WG-REM}}$} \\
\cmidrule(l{3pt}r{3pt}){3-4} \cmidrule(l{3pt}r{3pt}){5-6} \cmidrule(l{3pt}r{3pt}){7-8} \cmidrule(l{3pt}r{3pt}){9-10}
$\gamma_3$ & $\beta_3$ & Bias & SD & Bias & SD & Bias & SD & Bias & SD\\
\midrule
 & -2.0 & 1.356 & 0.323 & 0.032 & 0.089 & 0.110 & 0.123 & 0.004 & 0.094\\

\addlinespace[1pt]
 & -1.5 & 1.018 & 0.275 & 0.034 & 0.089 & 0.083 & 0.118 & 0.004 & 0.094\\

\addlinespace[1pt]
 & -1.0 & 0.680 & 0.235 & 0.032 & 0.089 & 0.056 & 0.114 & 0.004 & 0.094\\

\addlinespace[1pt]
 & -0.5 & 0.342 & 0.207 & 0.023 & 0.089 & 0.029 & 0.112 & 0.002 & 0.094\\

\addlinespace[1pt]
 & 0.5 & -0.334 & 0.208 & -0.016 & 0.088 & -0.026 & 0.114 & 0.000 & 0.094\\

\addlinespace[1pt]
 & 1.0 & -0.671 & 0.236 & -0.026 & 0.088 & -0.053 & 0.117 & -0.001 & 0.094\\

\addlinespace[1pt]
 & 1.5 & -1.009 & 0.276 & -0.028 & 0.088 & -0.080 & 0.122 & -0.002 & 0.094\\

\addlinespace[1pt]
\multirow{-8}{*}{\raggedleft\arraybackslash -2} & 2.0 & -1.347 & 0.324 & -0.026 & 0.088 & -0.107 & 0.128 & -0.002 & 0.094\\
\cmidrule{1-10}
\addlinespace[1pt]
 & -2.0 & 1.290 & 0.312 & 0.020 & 0.084 & 0.149 & 0.127 & 0.005 & 0.087\\

\addlinespace[1pt]
 & -1.5 & 0.969 & 0.255 & 0.023 & 0.084 & 0.113 & 0.116 & 0.005 & 0.087\\

\addlinespace[1pt]
 & -1.0 & 0.648 & 0.206 & 0.024 & 0.084 & 0.077 & 0.108 & 0.005 & 0.087\\

\addlinespace[1pt]
 & -0.5 & 0.327 & 0.170 & 0.019 & 0.083 & 0.041 & 0.103 & 0.004 & 0.087\\

\addlinespace[1pt]
 & 0.5 & -0.315 & 0.174 & -0.012 & 0.083 & -0.032 & 0.101 & 0.001 & 0.087\\

\addlinespace[1pt]
 & 1.0 & -0.636 & 0.213 & -0.018 & 0.083 & -0.068 & 0.106 & 0.000 & 0.087\\

\addlinespace[1pt]
 & 1.5 & -0.957 & 0.264 & -0.017 & 0.083 & -0.104 & 0.113 & 0.000 & 0.087\\

\addlinespace[1pt]
\multirow{-8}{*}{\raggedleft\arraybackslash -1} & 2.0 & -1.278 & 0.322 & -0.015 & 0.083 & -0.141 & 0.123 & 0.000 & 0.087\\
\cmidrule{1-10}
\addlinespace[1pt]
 & -2.0 & -1.264 & 0.317 & -0.011 & 0.085 & -0.135 & 0.126 & 0.002 & 0.087\\

\addlinespace[1pt]
 & -1.5 & -0.946 & 0.260 & -0.013 & 0.085 & -0.099 & 0.117 & 0.002 & 0.087\\

\addlinespace[1pt]
 & -1.0 & -0.627 & 0.210 & -0.014 & 0.085 & -0.063 & 0.109 & 0.002 & 0.087\\

\addlinespace[1pt]
 & -0.5 & -0.309 & 0.173 & -0.009 & 0.084 & -0.027 & 0.103 & 0.003 & 0.087\\

\addlinespace[1pt]
 & 0.5 & 0.328 & 0.174 & 0.022 & 0.084 & 0.044 & 0.101 & 0.006 & 0.087\\

\addlinespace[1pt]
 & 1.0 & 0.646 & 0.211 & 0.027 & 0.085 & 0.080 & 0.104 & 0.007 & 0.087\\

\addlinespace[1pt]
 & 1.5 & 0.964 & 0.261 & 0.027 & 0.085 & 0.116 & 0.110 & 0.007 & 0.087\\

\addlinespace[1pt]
\multirow{-8}{*}{\raggedleft\arraybackslash 1} & 2.0 & 1.282 & 0.318 & 0.024 & 0.085 & 0.152 & 0.118 & 0.007 & 0.087\\
\cmidrule{1-10}
\addlinespace[1pt]
 & -2.0 & -1.349 & 0.310 & -0.030 & 0.092 & -0.104 & 0.124 & -0.005 & 0.094\\

\addlinespace[1pt]
 & -1.5 & -1.008 & 0.265 & -0.031 & 0.092 & -0.078 & 0.118 & -0.005 & 0.094\\

\addlinespace[1pt]
 & -1.0 & -0.667 & 0.228 & -0.030 & 0.091 & -0.053 & 0.114 & -0.004 & 0.094\\

\addlinespace[1pt]
 & -0.5 & -0.327 & 0.203 & -0.020 & 0.091 & -0.027 & 0.111 & -0.003 & 0.094\\

\addlinespace[1pt]
 & 0.5 & 0.355 & 0.209 & 0.020 & 0.092 & 0.024 & 0.110 & -0.001 & 0.094\\

\addlinespace[1pt]
 & 1.0 & 0.695 & 0.239 & 0.029 & 0.092 & 0.049 & 0.112 & 0.000 & 0.094\\

\addlinespace[1pt]
 & 1.5 & 1.036 & 0.279 & 0.031 & 0.092 & 0.075 & 0.116 & 0.001 & 0.094\\

\addlinespace[1pt]
\multirow{-8}{*}{\raggedleft\arraybackslash 2} & 2.0 & 1.377 & 0.326 & 0.029 & 0.092 & 0.100 & 0.121 & 0.001 & 0.094\\
\bottomrule
\end{tabular}
\begin{tablenotes}
\item True ATE $\tau = 1$ in all settings. Clusters satisfying $p_j \notin (0.05,\, 0.95)$ were excluded to ensure adequate covariate overlap. Within-group estimators used $G = 5$ groups formed by k-means clustering on observed cluster-level treatment prevalence $p_j$.
\end{tablenotes}
\end{threeparttable}
\end{table}

\begin{table}[!h]
\centering
\caption{\label{tab:s1-cls}Simulation results for Scenario 1 (unmeasured cluster-level confounding only with $\theta = 0$): bias and standard deviation (SD) of the four g-computation estimators across combinations of number of clusters $J$ and cluster size $n_j$, with $\gamma_3 = \beta_3 = 1.5$ based on 500 Monte Carlo replications. $\tilde{\tau}_{\text{LM}}$, $\tilde{\tau}_{\text{WG-LM}}$, $\tilde{\tau}_{\text{REM}}$, and $\tilde{\tau}_{\text{WG-REM}}$ denote the standard linear model, within-group linear model, standard random-effects model, and within-group random-effects g-computation estimators, respectively.}
\centering
\fontsize{9}{11}\selectfont
\begin{threeparttable}
\begin{tabular}[t]{rrrrrrrrrr}
\toprule
\multicolumn{2}{c}{ } & \multicolumn{2}{c}{$\tilde{\tau}_{\text{LM}}$} & \multicolumn{2}{c}{$\tilde{\tau}_{\text{REM}}$} & \multicolumn{2}{c}{$\tilde{\tau}_{   ext{WG-LM}}$} & \multicolumn{2}{c}{$\tilde{\tau}_{text{WG-REM}}$} \\
\cmidrule(l{3pt}r{3pt}){3-4} \cmidrule(l{3pt}r{3pt}){5-6} \cmidrule(l{3pt}r{3pt}){7-8} \cmidrule(l{3pt}r{3pt}){9-10}
$J$ & $n_j$ & Bias & SD & Bias & SD & Bias & SD & Bias & SD\\
\midrule
 & 15 & 1.238 & 0.307 & 0.034 & 0.099 & 0.123 & 0.140 & 0.002 & 0.105\\

\addlinespace[2pt]
 & 30 & 1.221 & 0.276 & 0.019 & 0.068 & 0.099 & 0.100 & 0.002 & 0.070\\

\addlinespace[2pt]
 & 45 & 1.180 & 0.270 & 0.011 & 0.057 & 0.096 & 0.086 & -0.002 & 0.059\\

\addlinespace[2pt]
 & 60 & 1.189 & 0.258 & 0.011 & 0.048 & 0.092 & 0.080 & 0.001 & 0.051\\

\addlinespace[2pt]
 & 75 & 1.283 & 0.283 & 0.009 & 0.043 & 0.104 & 0.081 & 0.002 & 0.045\\

\addlinespace[2pt]
 & 90 & 1.288 & 0.286 & 0.008 & 0.041 & 0.104 & 0.077 & 0.002 & 0.043\\

\addlinespace[2pt]
 & 105 & 1.221 & 0.274 & 0.007 & 0.035 & 0.098 & 0.071 & 0.002 & 0.037\\

\addlinespace[2pt]
 & 120 & 1.238 & 0.272 & 0.008 & 0.035 & 0.096 & 0.069 & 0.004 & 0.036\\

\addlinespace[2pt]
 & 135 & 1.278 & 0.279 & 0.001 & 0.032 & 0.095 & 0.070 & -0.004 & 0.034\\

\addlinespace[2pt]
\multirow{-10}{*}{\raggedleft\arraybackslash 50} & 150 & 1.268 & 0.266 & 0.003 & 0.031 & 0.100 & 0.073 & 0.000 & 0.032\\
\cmidrule{1-10}
\addlinespace[2pt]
 & 15 & 1.257 & 0.210 & 0.040 & 0.069 & 0.154 & 0.098 & 0.007 & 0.075\\

\addlinespace[2pt]
 & 30 & 1.237 & 0.197 & 0.021 & 0.047 & 0.127 & 0.073 & 0.004 & 0.050\\

\addlinespace[2pt]
 & 45 & 1.228 & 0.196 & 0.011 & 0.040 & 0.112 & 0.065 & -0.001 & 0.043\\

\addlinespace[2pt]
 & 60 & 1.229 & 0.192 & 0.012 & 0.033 & 0.109 & 0.056 & 0.003 & 0.034\\

\addlinespace[2pt]
 & 75 & 1.306 & 0.181 & 0.008 & 0.030 & 0.116 & 0.057 & 0.001 & 0.032\\

\addlinespace[2pt]
 & 90 & 1.283 & 0.198 & 0.008 & 0.027 & 0.119 & 0.056 & 0.003 & 0.028\\

\addlinespace[2pt]
 & 105 & 1.269 & 0.194 & 0.004 & 0.026 & 0.108 & 0.053 & -0.002 & 0.027\\

\addlinespace[2pt]
 & 120 & 1.263 & 0.201 & 0.003 & 0.024 & 0.108 & 0.051 & -0.001 & 0.024\\

\addlinespace[2pt]
 & 135 & 1.305 & 0.189 & 0.004 & 0.021 & 0.116 & 0.050 & 0.000 & 0.023\\

\addlinespace[2pt]
\multirow{-10}{*}{\raggedleft\arraybackslash 100} & 150 & 1.289 & 0.190 & 0.004 & 0.021 & 0.110 & 0.051 & 0.000 & 0.022\\
\bottomrule
\end{tabular}
\begin{tablenotes}
\item True ATE $\tau = 1$ in all settings. Clusters satisfying $p_j \notin (0.05,\, 0.95)$ were excluded to ensure adequate covariate overlap. Within-group estimators used $G = 5$ groups formed by k-means clustering on observed cluster-level treatment prevalence $p_j$.
\end{tablenotes}
\end{threeparttable}
\end{table}

\begin{table}[!h]
\centering
\caption{\label{tab:s2-theta}Simulation results for Scenario 2 (unmeasured cluster-level confounding and treatment effect heterogeneity): bias and standard deviation (SD) of the four g-computation estimators across values of the effect heterogeneity parameter $\theta$, with $J = 50$ clusters, $n_j = 20$ individuals per cluster, and $\gamma_3 = \beta_3 = 1.5$, based on 500 Monte Carlo replications. $\tilde{\tau}_{\text{LM}}$, $\tilde{\tau}_{\text{WG-LM}}$, $\tilde{\tau}_{\text{REM}}$, and $\tilde{\tau}_{\text{WG-REM}}$ denote the standard linear model, within-group linear model, standard random-effects model, and within-group random-effects g-computation estimators, respectively.}
\centering
\fontsize{9}{11}\selectfont
\begin{threeparttable}
\begin{tabular}[t]{rrrrrrrrr}
\toprule
\multicolumn{1}{c}{ } & \multicolumn{2}{c}{$\tilde{\tau}_{\text{LM}}$} & \multicolumn{2}{c}{$\tilde{\tau}_{\text{REM}}$} & \multicolumn{2}{c}{$\tilde{\tau}_{\text{WG-LM}}$} & \multicolumn{2}{c}{$\tilde{\tau}_{\text{WG-REM}}$} \\
\cmidrule(l{3pt}r{3pt}){2-3} \cmidrule(l{3pt}r{3pt}){4-5} \cmidrule(l{3pt}r{3pt}){6-7} \cmidrule(l{3pt}r{3pt}){8-9}
$\theta$ & Bias & SD & Bias & SD & Bias & SD & Bias & SD\\
\midrule
-1.5 & 1.408 & 0.386 & 0.582 & 0.264 & 0.516 & 0.272 & 0.454 & 0.260\\
\addlinespace[2pt]
-1.3 & 1.363 & 0.358 & 0.508 & 0.233 & 0.460 & 0.242 & 0.393 & 0.229\\
\addlinespace[2pt]
-1.1 & 1.317 & 0.333 & 0.435 & 0.203 & 0.404 & 0.214 & 0.333 & 0.200\\
\addlinespace[2pt]
-0.9 & 1.272 & 0.312 & 0.361 & 0.175 & 0.349 & 0.186 & 0.274 & 0.173\\
\addlinespace[2pt]
-0.7 & 1.227 & 0.296 & 0.287 & 0.148 & 0.293 & 0.161 & 0.214 & 0.147\\
\addlinespace[2pt]
-0.5 & 1.181 & 0.284 & 0.214 & 0.123 & 0.237 & 0.139 & 0.155 & 0.124\\
\bottomrule
\end{tabular}
\begin{tablenotes}
\item True ATE $\tau = 1$ in all settings. The individual treatment effect is $Y_{ij}(1) - Y_{ij}(0) = \zeta + \theta U_j^2$, where $zeta = 1 - \theta\, \mathbb{E}(U_j^2)$ ensures $\tau = 1$ across all $\theta$. Clusters satisfying $p_j \notin (0.05,\, 0.95)$ were excluded to ensure adequate covariate overlap. Within-group estimators used $G = 5$ groups formed by k-means clustering on observed cluster-level treatment prevalence $p_j$.
\end{tablenotes}
\end{threeparttable}
\end{table}

\begin{table}[!h]
\centering
\caption{\label{tab:s2-cls}Simulation results for Scenario 2 (unmeasured cluster-level confounding and treatment effect heterogeneity): bias and standard deviation (SD) of the four g-computation estimators across combinations of number of clusters $J$ and cluster size $n_j$, with $\gamma_3 = \beta_3 = 1.5$ and $\theta = -1$, based on 500 Monte Carlo replications. $\tilde{\tau}_{\text{LM}}$, $\tilde{\tau}_{\text{WG-LM}}$, $\tilde{\tau}_{\text{REM}}$, and $\tilde{\tau}_{\text{WG-REM}}$ denote the standard linear model, within-group linear model, standard random-effects model, and within-group random-effects g-computation estimators, respectively.}
\centering
\fontsize{9}{11}\selectfont
\begin{threeparttable}
\begin{tabular}[t]{rrrrrrrrrr}
\toprule
\multicolumn{2}{c}{ } & \multicolumn{2}{c}{$\tilde{\tau}_{\text{LM}}$} & \multicolumn{2}{c}{$\tilde{\tau}_{\text{REM}}$} & \multicolumn{2}{c}{$\tilde{\tau}_{\text{WG-LM}}$} & \multicolumn{2}{c}{$\tilde{\tau}_{\text{WG-REM}}$} \\
\cmidrule(l{3pt}r{3pt}){3-4} \cmidrule(l{3pt}r{3pt}){5-6} \cmidrule(l{3pt}r{3pt}){7-8} \cmidrule(l{3pt}r{3pt}){9-10}
$J$ & $n_j$ & Bias & SD & Bias & SD & Bias & SD & Bias & SD\\
\midrule
 & 15 & 1.377 & 0.343 & 0.366 & 0.182 & 0.318 & 0.206 & 0.227 & 0.177\\

\addlinespace[2pt]
 & 30 & 1.377 & 0.324 & 0.367 & 0.165 & 0.323 & 0.179 & 0.256 & 0.161\\

\addlinespace[2pt]
 & 45 & 1.345 & 0.305 & 0.357 & 0.159 & 0.328 & 0.171 & 0.256 & 0.158\\

\addlinespace[2pt]
 & 60 & 1.349 & 0.325 & 0.359 & 0.164 & 0.322 & 0.168 & 0.258 & 0.155\\

\addlinespace[2pt]
 & 75 & 1.414 & 0.331 & 0.352 & 0.161 & 0.306 & 0.166 & 0.237 & 0.152\\

\addlinespace[2pt]
 & 90 & 1.421 & 0.329 & 0.360 & 0.149 & 0.316 & 0.165 & 0.247 & 0.147\\

\addlinespace[2pt]
 & 105 & 1.369 & 0.313 & 0.361 & 0.157 & 0.324 & 0.161 & 0.258 & 0.152\\

\addlinespace[2pt]
 & 120 & 1.388 & 0.320 & 0.362 & 0.151 & 0.325 & 0.159 & 0.258 & 0.147\\

\addlinespace[2pt]
 & 135 & 1.412 & 0.312 & 0.356 & 0.161 & 0.309 & 0.164 & 0.243 & 0.149\\

\addlinespace[2pt]
\multirow{-10}{*}{\raggedleft\arraybackslash 50} & 150 & 1.410 & 0.319 & 0.356 & 0.152 & 0.315 & 0.162 & 0.247 & 0.146\\
\cmidrule{1-10}
\addlinespace[2pt]
 & 15 & 1.373 & 0.239 & 0.370 & 0.125 & 0.339 & 0.146 & 0.236 & 0.125\\

\addlinespace[2pt]
 & 30 & 1.389 & 0.224 & 0.371 & 0.117 & 0.348 & 0.126 & 0.261 & 0.112\\

\addlinespace[2pt]
 & 45 & 1.384 & 0.228 & 0.368 & 0.112 & 0.341 & 0.123 & 0.263 & 0.111\\

\addlinespace[2pt]
 & 60 & 1.399 & 0.221 & 0.381 & 0.109 & 0.353 & 0.115 & 0.279 & 0.106\\

\addlinespace[2pt]
 & 75 & 1.438 & 0.220 & 0.366 & 0.107 & 0.322 & 0.113 & 0.247 & 0.103\\

\addlinespace[2pt]
 & 90 & 1.420 & 0.220 & 0.359 & 0.105 & 0.332 & 0.117 & 0.253 & 0.103\\

\addlinespace[2pt]
 & 105 & 1.410 & 0.224 & 0.357 & 0.109 & 0.328 & 0.119 & 0.251 & 0.107\\

\addlinespace[2pt]
 & 120 & 1.413 & 0.232 & 0.357 & 0.112 & 0.331 & 0.117 & 0.256 & 0.106\\

\addlinespace[2pt]
 & 135 & 1.427 & 0.229 & 0.353 & 0.115 & 0.324 & 0.120 & 0.247 & 0.108\\

\addlinespace[2pt]
\multirow{-10}{*}{\raggedleft\arraybackslash 100} & 150 & 1.425 & 0.233 & 0.360 & 0.109 & 0.323 & 0.117 & 0.251 & 0.104\\
\bottomrule
\end{tabular}
\begin{tablenotes}
\item True ATE $\tau = 1$ in all settings. The individual treatment effect is $Y_{ij}(1) - Y_{ij}(0) = \zeta + \theta U_j^2$, where $\zeta = 1 - \theta\, \mathbb{E}(U_j^2)$ ensures $\tau = 1$ across all $theta$. Clusters satisfying $p_j \notin (0.05,\, 0.95)$ were excluded to ensure adequate covariate overlap. Within-group estimators used $G = 5$ groups formed by k-means clustering on observed cluster-level treatment prevalence $p_j$.
\end{tablenotes}
\end{threeparttable}
\end{table}

\begin{figure}[!ht]
\centering
\includegraphics[width=0.6\textwidth]{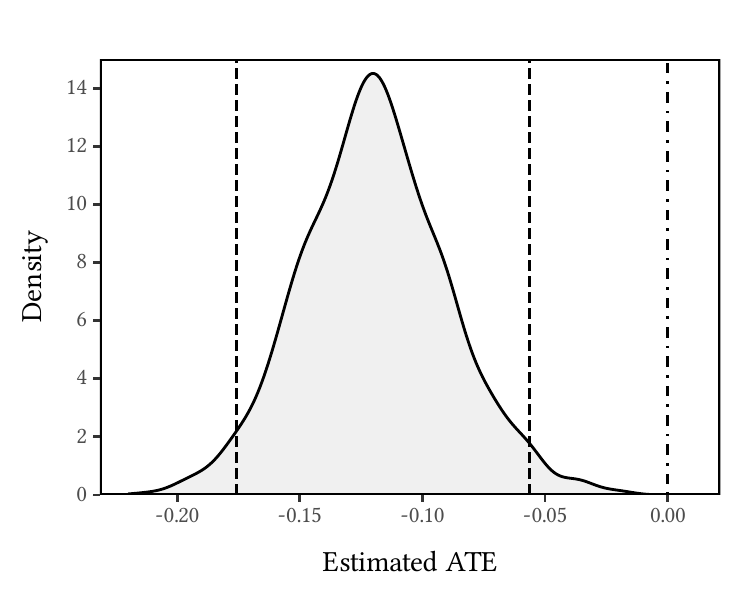}
\caption{Distribution of the within-group REM-based g-computation ATE estimates
across 1,000 bootstrap replications. The vertical dotted line represents the
null effect; dashed lines indicate the 2.5th and 97.5th percentiles,
corresponding to the empirical 95\% confidence interval.}
\label{fig:boot-hist-plot}
\end{figure}

\end{document}